\documentclass[journal=nalefd,manuscript=letter, layout = twocolumn ]{achemso}

\usepackage[version=3]{mhchem} % Formula subscripts using \ce{}
\usepackage{xcolor}
\usepackage{soul}
\usepackage{natbib}
\usepackage{mciteplus}
\usepackage{multicol,graphicx,float}
\usepackage{comment}
\usepackage{placeins}
\usepackage[separate-uncertainty = true,multi-part-units=single]{siunitx}
\usepackage{amsmath,amssymb}
\newcommand*\mycommand[1]{\texttt{\emph{#1}}}

\author{Leonard Geilen}
\affiliation{Walter Schottky Institute and Physics Department, Technical University of Munich, 85748 Garching, Germany.}
\author{Lukas Schleicher}
\affiliation{Department of Electrical Engineering, School of Computation, Information and Technology, Technical University of Munich, 85748 Garching, Germany.}
\author{Alexander Musta}
\affiliation{Walter Schottky Institute and Physics Department, Technical University of Munich, 85748 Garching, Germany.}
\author{Benedict Brouwer}
\affiliation{Walter Schottky Institute and Physics Department, Technical University of Munich, 85748 Garching, Germany.}
\author{Eva M. Weig}
\affiliation{Department of Electrical Engineering, School of Computation, Information and Technology, Technical University of Munich, 85748 Garching, Germany.}
\alsoaffiliation{Munich Center for Quantum Science and Technology (MCQST), Schellingstr. 4, 80799 Munich, Germany.}
\alsoaffiliation{TUM Center for Quantum Engineering (ZQE), 85748 Garching, Germany.}
\author{Alexander Holleitner}
\affiliation{Walter Schottky Institute and Physics Department, Technical University of Munich, 85748 Garching, Germany.}
\alsoaffiliation{Munich Center for Quantum Science and Technology (MCQST), Schellingstr. 4, 80799 Munich, Germany.}
\author{Anne Rodriguez}
\affiliation{Department of Electrical Engineering, School of Computation, Information and Technology, Technical University of Munich, 85748 Garching, Germany.}
\alsoaffiliation{Munich Center for Quantum Science and Technology (MCQST), Schellingstr. 4, 80799 Munich, Germany.}
\email{holleitner@wsi.tum.de, eva.weig@tum.de, anne.rodriguez@tum.de}

\title[An \textsf{achemso} demo]
  {In-situ profiling of pressure-induced exciton traps in suspended MoS$_2$ monolayers}

\begin{document}

% \begin{tocentry}

% Some journals require a graphical entry for the Table of Contents.
% This should be laid out ``print ready'' so that the sizing of the

% Inside the \texttt{tocentry} environment, the font used is Helvetica
% 8\,pt, as required by \emph{Journal of the American Chemical

% The surrounding frame is 9\,cm by 3.5\,cm, which is the maximum
% permitted for  \emph{Journal of the American Chemical Society}
% graphical table of content entries. The box will not resize if the
% content is too big: instead it will overflow the edge of the box.

% This box and the associated title will always be printed on a
% separate page at the end of the document.

% \end{tocentry}

\begin{abstract}
We demonstrate the in-situ read-out of the spatial profile of suspended MoS$_2$ monolayers hosted on substrates with nano-structured holes. As the  profiles are spatially bent, the suspended MoS$_2$ monolayers act as exciton traps with tunable luminescence intensity and energy. The tunability is realized by controlling the environmental pressure on the monolayers, which allows to control hundreds of suspended MoS$_2$ monolayers on a single substrate.  The in-situ read-out is based on Fabry-Pérot interferences and a model of the corresponding reflectance contrast maps of the investigated monolayers. 
 
\end{abstract}

\section{Main text}
Exciton traps have been realized in several semiconductors and their heterostructures with the main goal to locally control and manipulate the light-matter interaction. Such traps allow achieving high exciton densities to study possible bosonic and fermionic many-body interactions particularly at low temperatures, \cite{high_condensation_2012, schinner_confinement_2013, alloing_optically_2013, cohen_dark_2016, combescot_boseeinstein_2017, shanks_nanoscale_2021, thureja_electrically_2022, katzer_exciton-phonon_2023, kumar_strain_2024, joe_controlled_2024} and for investigating exciton and polariton dynamics in photonic cavities and/or polariton devices \cite{balili_bose-einstein_2007, schneider_exciton-polariton_2016}. Laterally patterned potential landscapes further allow defining low-dimensional exciton circuits, where the exciton diffusion and drift dynamics as well as possible exciton transistor applications can be explored   \cite{hagn_electric-field-induced_1995, butov_macroscopically_2002, voros_long-distance_2005, gartner_drift_2006, fowler-gerace_voltage-controlled_2021, sun_excitonic_2022, thureja_electrically_2022, harats_dynamics_2020, lee_drift-dominant_2022}. The tunability of exciton traps typically relies on the Stark effect in electrostatic field-effect devices \cite{schinner_electrostatically_2011, thureja_electrically_2022, joe_controlled_2024} or on a local strain engineering \cite{sinclair_strain-induced_2011}. The latter can be introduced, e.g., by utilizing the tip of a needle pressing against the investigated materials \cite{sinclair_strain-induced_2011} or by electro-static tuning of certain strain components in suspended semiconductor membranes\cite{kumar_strain_2024}. Similarly, nanoscale pillars and pyramids can imprint spatial profiles onto thin materials, such as transition-metal dichalcogenides (TMDCs), giving rise to local strain variations and, therefore, excitonic potential landscapes\cite{branny_deterministic_2017, blundo_evidence2_2020, parto_defect_2021, rosati_dark_2021}. In all cases, it is highly desirable to read out the spatial profile to quantify the strain profiles of the investigated exciton traps, while performing the optical experiments on the exciton dynamics. 

Here, we demonstrate a pressure-induced exciton trapping scheme in suspended MoS$_2$ monolayers exfoliated onto nano-structured and metallized substrates exhibiting circular holes. In a first step, we extract the spatial profiles of the suspended MoS$_2$ monolayers by atomic force microscopy and successfully describe the profiles by the so-called extended Hencky model for pressure-strained membranes\cite{lloyd_band_2016, ma_extended_2018}. Then, we show that spatially resolved optical reflectance measurements allow us to extract the spatial deflection profile of the suspended MoS$_2$ monolayers in an in-situ manner. This alternative method allows us to access the spatial profile of the MoS$_2$ monolayer investigated under optical illumination and under real-time pressure and excitation conditions. We introduce a dielectric model to describe the corresponding Fabry-Pérot-type interferences, which evolve between the suspended, spatially-bent MoS$_2$ monolayers and the reflective coating at the bottom of the nanostructured holes. We highlight that controlling the environmental pressure of the nanostructured chip allows tuning the spatial profile of the MoS$_2$ monolayers and, therefore, the exciton trapping potentials in a reversible manner. Since the demonstrated scheme allows fabricating hundreds of exciton traps on a substrate in a scalable way, our results open the pathway for photonic circuits with excitons trapped in suspended TMDCs and their respective van-der Waals heterostructures.

\begin{figure}[ht!]
    \centering
    \includegraphics[]{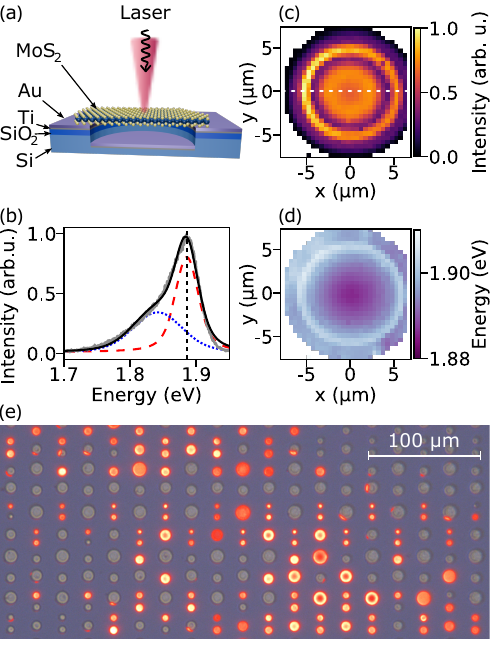}
    \caption{Schematic of the exciton trap in suspended monolayers of MoS$_2$. (a) A MoS$_2$ monolayer deposited on a patterned Si/SiO$_2$ substrate coated with Ti/Au is illuminated by a laser. (b) Typical photoluminescence (PL) spectrum measured at the center of a suspended MoS$_2$ monolayer (gray data), fitted with the sum of two Voigt functions (solid line) representing contributions from excitons (dashed line) as well as trions and defects (dotted line), respectively. Excitonic PL (c) intensity and (d) energy maps of the same suspended MoS$_2$ monolayer as in (b) with diameter of \SI{13}{\micro\meter}. (e) Optical microscope image of a sample with various suspended MoS$_2$ layers overlayed with their corresponding photoluminescence (red).}
    \label{fig:fig1}
\end{figure}
Figure \ref{fig:fig1}a sketches a suspended MoS$_2$ monolayer atop a pre-structured substrate, which consist of Si/SiO$_2$ (\SI{500}{\micro\meter}/\SI{285}{\nano\meter}). 
Circular holes are lithographically defined and dry-etched \SI{2}{\micro\meter} deep into the substrate. Then, the chip is metallized with a thin layer of Ti/Au (\SI{2}{\nano\meter}/\SI{5}{\nano\meter}).
The MoS$_2$ is stacked onto the Ti/Au using the gold-assisted direct exfoliation method with an annealing temperature of \SI{200}{\degreeCelsius} at ambient conditions~\cite{velicky_mechanism-of-gold_2018}. The resulting circularly shaped, suspended MoS$_2$ layers have diameters between \SI{3}{\micro\meter} and \SI{14}{\micro\meter}, and we particularly concentrate on MoS$_2$ monolayers. We excite them with a cw-laser at a photon energy of  \SI{2.33}{eV} (\SI{532}{nm}) and measure the corresponding photoluminescence (PL) at room temperature. Figure \ref{fig:fig1}b shows a PL spectrum measured at the center of one of the suspended MoS$_2$ monolayers. We fit the PL spectrum with the sum of two Voigt functions to extract the $X^0_\text{1S}$ exciton emission energy to be \SI{1.89}{eV} (vertical dashed line) in combination with a tail towards lower energies representing the luminescence stemming from trions and/or defects.\cite{chow_defect-induced_2015, Klein_robust_2018} Figures \ref{fig:fig1}c and d show the spatial dependence of the excitonic PL of the same circularly shaped, suspended MoS$_2$ monolayer. Importantly, the overall PL intensity depicted in Fig.~\ref{fig:fig1}c has a rather broad maximum at the center of the suspended MoS$_2$ monolayer. 
\begin{figure}[ht!]
    \centering
    \includegraphics[scale=.85]{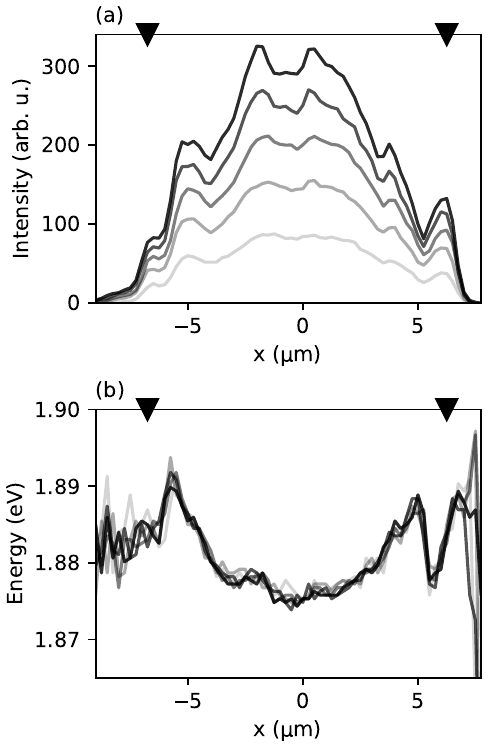}
    \caption{Luminescence profile across a suspended MoS$_2$ monolayer. (a) PL maximum across the center of a suspended MoS$_2$ monolayer with a diameter of \SI{13}{\micro\meter} for five equidistant laser powers ranging from \SI{1}{\micro\watt} (gray) to \SI{5}{\micro\watt} (black). Black triangles indicate edges of the suspended part of the MoS$_2$ monolayer. (b) Energy profile of the photon emission as in (a). } 
    \label{fig:fig2}
\end{figure}
We note that the peripheral maxima at the edge can be explained by an interference effect, as discussed below. The central intensity maximum is associated with a minimum in emission energy as depicted in Fig.~\ref{fig:fig1}d. Generally, the direct exfoliation method allows us to transfer mono- and few-layers of MoS$_2$ on a large area of the pre-patterned and metallized substrates. Exemplarily, Fig.~\ref{fig:fig1}e depicts an optical microscope image of one of the investigated samples overlayed by a PL map filtered at the exciton photon energy (cf. dashed vertical line in Fig.~\ref{fig:fig1}b). The figure reveals a large number of circular luminescence patterns corresponding to circularly shaped areas of suspended MoS$_2$ layers with the brightest ones being identified as monolayers (cf. Supporting Information)~\cite{mak_atomically_2010}.

\begin{figure}[ht!]
    \centering
    \includegraphics[]{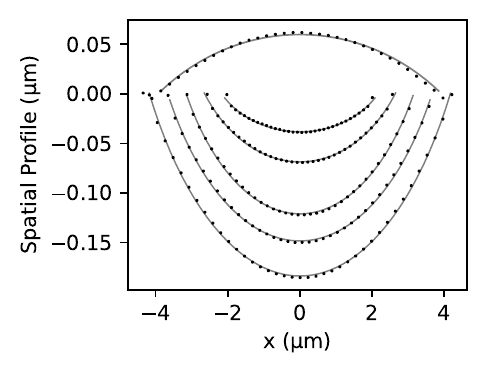}
    \caption{Atomic force microscope (AFM) line-cuts (dots) across suspended MoS$_2$ monolayers with diameters ranging from \SI{4}{\micro\meter} to \SI{8}{\micro\meter}. Solid lines represent fits with the extended Hencky model for pressure-strained membranes.}
    \label{fig:fig3}
\end{figure}

Figure \ref{fig:fig2}a shows the PL profile across one of the investigated suspended MoS$_2$ monolayers for various excitation powers. We observe that the excitonic photon emission is maximum at the center for all investigated powers, suggesting that most excitons are generated and trapped at the center of the suspended MoS$_2$ (see discussion below). Consistently, the energy of the corresponding photon emission exhibits a minimum at the center of the suspended MoS$_2$ for all powers (Fig.~\ref{fig:fig2}b). The triangles on the top indicate the edges of the investigated suspended MoS$_2$ monolayer with a diameter of \SI{13}{\micro\meter}.
\begin{figure}[p]
    \centering
    \includegraphics[]{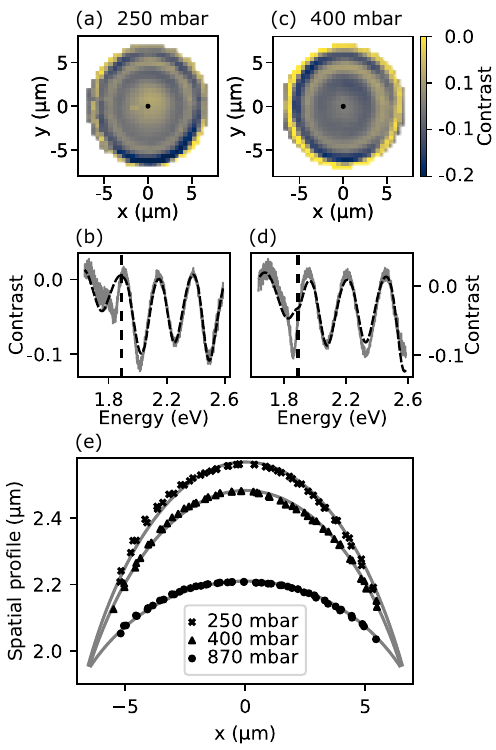}
\caption{In-situ reflectance characterization and determination of the spatial deflection of suspended MoS$_2$ monolayers. (a) Reflectance contrast map of a suspended MoS$_2$ monolayer with a diameter of \SI{13}{\micro\meter} at a wavelength of \SI{656}{\nano\meter} ($E_\text{photon}$ = 1.89  eV). The external pressure is set to be $p_\text{ext}$ = \SI{250\pm30}{mbar}. (b) Reflectance spectrum measured at the center of the suspended MoS$_2$ [cf. central dot in (a)]. Data (gray) are fitted by a reflectance model based on literature values of the relevant dielectric functions (dashed line). Dashed vertical line indicates the luminescence emission energy of the $X^0_\text{1S}$ exciton as in Fig.~\ref{fig:fig1}b, which is also the photon energy for the reflectance map in (a). (c) and (d): Same measurements and fit as in (a) and (b) for $p_\text{ext}$ = \SI{400\pm30}{mbar}.  (e) Deduced spatial deflection for $p_\text{ext}$ = \SI{250}{mbar} (crosses), \SI{400}{mbar} (triangles), and \SI{870}{mbar} (circles). Lines are fits to the data based on the extended Hencky model for pressure-strained membranes.}
    \label{fig:fig4}
\end{figure}

We characterize the topography of the suspended MoS$_2$ monolayers by means of an atomic force microscope (AFM).  Figure~\ref{fig:fig3} depicts corresponding AFM line-cuts through the center of several suspended monolayers with varying diameters (black dots). The lines are fits to the data following the extended Hencky model (cf. Supporting Information), which is typically used to compute the spatial deflection of pressure-strained membranes\cite{lloyd_band_2016, ma_extended_2018}. The lines fit the data very well for all investigated MoS$_2$ monolayers with the given diameters, suggesting that there exists a pressure difference of about $\Delta p = p_\text{int} - p_\text{ext} =$ \qtyrange[range-units=single,range-phrase=~--~]{10}{60}{mbar} between the enclosed volume below the suspended MoS$_2$ monolayers (with pressure $p_\text{int}$) and the outside ($p_\text{ext}$). 
We point out that most of the MoS$_2$ monolayers exhibit a negative deflection, which we explain by an increased temperature during the fabrication of the samples in combination with diffusion of the enclosed gases \cite{bunch_impermeable_2008}. Notably, there are also MoS$_2$ monolayers with a positive deflection even at ambient conditions (top line), which highlight that the MoS$_2$ tightly seals the enclosed volumes even in the case of overpressure~\cite{liu_atomically_2014}.

Generally, the pressure difference $\Delta p$ can slowly equilibrate over time and is subject to change with varying ambient (pressure) conditions. Therefore, it is essential to know the specific deflection profile of the investigated MoS$_2$ monolayers while performing optical experiments. We demonstrate that the deflection can be deduced in-situ from reflectance measurements.  Figure \ref{fig:fig4}a shows a reflectance map of a particular MoS$_2$ monolayer with a diameter of \SI{13}{\micro\meter} with an unknown internal pressure, placed inside of a vacuum chamber providing a nominal external pressure of $p_\text{ext} =$ \SI{250 \pm 30}{mbar}. The reflectance map is shown for $E_\text{photon}$ = \SI{1.89}{eV}, which coincides with the excitonic emission energy. One can clearly observe a ring-structure of the reflectance contrast as in the radial luminescence maps (cf. Fig.~\ref{fig:fig1}c). The reflectance maxima and minima depend both, on the position within the plane of the suspended monolayer and on the particular photon energy, at which the reflectance is measured. To demonstrate this Fabry-Pérot-type interference, we plot the reflectance contrast spectrum for a center position (gray data in Fig.~\ref{fig:fig4}b). Based on the literature values of the dielectric function of unstrained MoS$_2$ monolayers and the ones of the substrate materials (cf. sketch in Fig.~\ref{fig:fig1}a), we calculate the reflectance spectra utilizing the transfer matrix method (dashed line in Fig.~\ref{fig:fig4}b, cf.  Supporting Information for details)~\cite{Sigl_optical_2022}. The model accurately captures the number of reflectance minima and maxima throughout the whole investigated spectrum. We tentatively explain the deviation of the model to the data at a photon energy of $E_\text{photon}$\SI{\sim 1.89}{eV} by the impact of strain on the exciton and trion contributions within the dielectric function of MoS$_2$ (vertical dashed line highlights the energy of the $X^0_\text{1S}$ exciton as in Fig.~\ref{fig:fig1}b)~\cite{conley_bandgap-engineering_2013, kumar_strain_2024}. For $p_\text{ext} =$ \SI{400\pm30}{mbar}, the reflectance contrast map again displays the overall ring-shaped structure (Fig.~\ref{fig:fig4}c), but on trend, we observe fewer minima and maxima in the reflectance spectra (e.g. Fig.~\ref{fig:fig4}d for central position). Again, the model captures the number of reflectance minima and maxima very well (dashed line). 
As another output, the model gives an estimate of the height difference between the MoS$_2$ monolayer and the reflective metal bottom of the etched hole for each position on the monolayer, which in a second step, allows us to reconstruct the spatial profile of the suspended MoS$_2$ monolayer. Figure~\ref{fig:fig4}e shows corresponding line-cuts through such a two-dimensional map for three values of $p_\text{ext}$ (\SI{250}{mbar}, \SI{400}{mbar}, and \SI{870}{mbar}). Importantly, we can fit the reconstructed profile again by the extended Hencky model for pressure-strained membranes (solid lines in Fig.~\ref{fig:fig4}e as for the fitted AFM line-cuts in Fig.~\ref{fig:fig3}, cf. Supporting Information), suggesting that this particular MoS$_2$ monolayer is bent outwards at the investigated values of $p_\text{ext}$. We note that the value of the computed $p_\text{int}$ depends on the Young's modulus of the specific MoS$_2$ monolayers. 
%
% Here we use a value of $E=\SI{270}{\giga\pascal}$ deduced from the AFM measurements. While this value is in line with the literature~\cite{li_mapping_2018,bertolazzi_stretching_2011} demonstrating values between $\SI{265}{\giga\pascal}$ and $\SI{270}{\giga\pascal}$, it is associated with a rather large uncertainty such that the precise determination of the Young's modulus remains beyond the scope of the current manuscript. 
%
Here we use a value of $E=\SI{270}{\giga\pascal}$ as reported in literature~\cite{li_mapping_2018,bertolazzi_stretching_2011}. Note that the value obtained from the AFM measurements is associated with a rather large uncertainty such that the determination of the Young's modulus remains beyond the scope of the current manuscript. 
Correspondingly, the determined $p_\text{int}$ is subject to a certain scaling with $E$. However, we observe that to first order, the product $p_\text{int}~\cdot V_\text{enclosed}$ is constant within a set of experiments with varying $p_\text{ext}$ and $V_\text{enclosed}$ the computed volume underneath the deflected MoS$_2$, as expected for an enclosed ideal gas. 
\FloatBarrier
\begin{figure}[ht!]
    \centering
    % Figure needs to be adapted to width
    \includegraphics[]{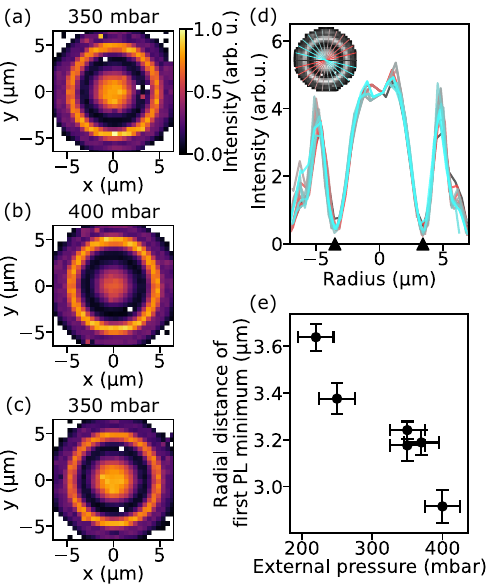}
    \caption{Reproducible tuning of the exciton trap with external pressure $p_\text{ext}$. (a) PL intensity map of a suspended MoS$_2$ monolayer with diameter of 13 µm at $p_\text{ext}$ = \SI{350}{mbar}, (b) \SI{400}{mbar}, and (c) \SI{350}{mbar}. In each case, the experimental uncertainty is \SI{\pm30}{mbar}. (d) Radial cross sections for different azimuthal angles with colors defined in the inset. The triangles highlight the first PL-minimum with respect to the center of the MoS$_2$. (e) Radial distance of the first PL-minimum (triangles in d) with respect to the center of the suspended MoS$_2$ monolayer as a function of $p_\text{ext}$.
    }
    \label{fig:fig5}
\end{figure}

In a next step, we demonstrate that the luminescence of the suspended MoS$_2$ monolayers can be manipulated in a reproducible way when changing $p_\text{ext}$.  Figure \ref{fig:fig5}a depicts the luminescence map of the same monolayer as in Fig.~\ref{fig:fig4} at  $p$$_\text{ext}$ = \SI{350\pm30}{mbar} with a clear luminescence maximum at the center.  When increasing $p$$_\text{ext}$ to \SI{400\pm30}{mbar}, we observe that the intensity and the radius of the inner luminescence maximum are reduced (cf. Fig.~\ref{fig:fig5}b), but both can be recovered when the external pressure is put to its initial value of $p$$_\text{ext}$ = \SI{350\pm30}{mbar} (cf. Fig.~\ref{fig:fig5}c). To describe the pressure-induced changes of the luminescence more quantitatively, Fig.~\ref{fig:fig5}d depicts radial cuts as illustrated in the inset. We determine the radial distance of the first PL-minimum with respect to the center of the suspended MoS$_2$ monolayers for various azimuthal angles (inset of Fig.~\ref{fig:fig5}d). We repeat the procedure for various values of $p$$_\text{ext}$. Figure \ref{fig:fig5}e depicts the deduced positions, revealing a clear decrease for higher values of $p$$_\text{ext}$ and that all values for the measurements at \SI{350\pm30}{mbar} overlap within the given uncertainties. Both observations support the interpretation that for this particular membrane, the MoS$_2$ monolayer is bent outwards and that it can be reversibly manipulated by $p_\text{ext}$.  

Generally, the spatial luminescence profiles as in Figs.~\ref{fig:fig1}c and \ref{fig:fig2}a demonstrate that the suspended MoS$_2$ monolayers can be tuned to a configuration with the PL-intensity being maximum at the center, while the corresponding photon energy is minimum (Figs.~\ref{fig:fig1}d and \ref{fig:fig2}b), as requested for exciton traps  \cite{high_condensation_2012, schinner_confinement_2013, alloing_optically_2013, cohen_dark_2016, combescot_boseeinstein_2017, shanks_nanoscale_2021, thureja_electrically_2022, katzer_exciton-phonon_2023, kumar_strain_2024, joe_controlled_2024, schinner_electrostatically_2011, sinclair_strain-induced_2011, blundo_evidence1_2020}.  
Both, radial and azimuthal strain components seem to exist in the suspended MoS$_2$ monolayers, as deduced from the extended Hencky model, which consistently fit the AFM line-cuts (Fig.~\ref{fig:fig3}) as well as the spatial deflection derived from the reflectance measurements (Fig.~\ref{fig:fig4}e and Supporting Information). From the fits, we compute radial and azimuthal strain components with maximum values of approximately $0.3\%$ at the center of the suspended monolayers (cf. Supporting Information); with the general trend that monolayers with a larger diameter exhibit a larger maximum strain~\cite{lloyd_band_2016}.  All extracted maxima are below the value, where one expects a transition to indirect bandgap characteristics in MoS$_2$ monolayer~\cite{conley_bandgap-engineering_2013, blundo_evidence1_2020}. In other words, the PL can be interpreted to stem mainly from $X^0_\text{1S}$ excitons at the K / K'- points of the Brillouin zone~\cite{kumar_strain_2024, conley_bandgap-engineering_2013, rosati_dark_2021}.   Moreover, at the center of the suspended MoS$_2$ monolayers, the PL can comprise contributions from trions (e.g. Fig.~\ref{fig:fig1}c and Supporting Information). We interpret them to form between locally generated excitons and free charge carriers \cite{harats_dynamics_2020, lee_drift-dominant_2022}. The latter are reported to drift in TMDCs according to strain profiles even at room temperature, while the exciton drift plays a role only at low temperatures at the given dimensions \cite{harats_dynamics_2020, lee_drift-dominant_2022}. Based on the latter argument, we interpret the central PL maximum, as in Fig.~\ref{fig:fig2}a, to stem from an increased optical absorption based on the specific bending of the suspended MoS$_2$ monolayers in combination with the Fabry-Pérot type interference at the given laser excitation (cf. Fig.~\ref{fig:fig4} and Supporting Information).  
At the edge of the suspended monolayers (cf. triangles in Fig.~\ref{fig:fig2}), further effects and parameters, such as the roughness of the pre-etched and metallized rim, come into play for both, the specific spatial profile of the suspended MoS$_2$ monolayers and the illumination as well as the PL-intensity because of possible local variation in material texture, adhesion, adsorbed molecules, light reflection, and exciton quenching effects. 
Last but not least, we note that our results open the pathway for exciton traps in suspended TMDC monolayers with quantum emitters positioned centrally in the Fabry-Pérot cavity for possible quantum transduction schemes~\cite{kim_position_2019, gao_high-speed_2020}. 

In conclusion, we demonstrate that suspended MoS$_2$ monolayers with an enclosed gas volume allow to trap excitons at their center and that the spatial profile of the monolayers can be read out in-situ by measuring the optical reflectance of the spatially bent monolayers. The deduced profiles are  numerically described by the extended Hencky model and are consistent with line-cuts measured by an atomic force microscope. The luminescence intensity and energy are reversibly tunable by the external environmental pressure, such that hundreds of such pressure-induced exciton traps can be controlled at the same time.

\FloatBarrier

\begin{acknowledgement}
We gratefully acknowledge financial support by the Deutsche Forschungsgemeinschaft (DFG) via the Munich Center for Quantum Science and Technology (MCQST) - EXC 2111-390814868 and e-conversion - EXC 2089/1-390776260. L.G. and A.H. acknowledge support through the TUM International Graduate School of Science and Engineering (IGSSE), A.H and E.W. from the One Munich Strategy Forum - EQAP, and A.H. from the Munich Quantum Valley (MQV) K6.
\end{acknowledgement}

\bibliography{bibliography}

\end{document}